\documentclass[preprint,showpacs,amsmath,amssymb]{revtex4}
\usepackage{mathrsfs}

\usepackage{graphicx,color}
\usepackage{dcolumn}
\usepackage{bm}

\begin{document}

\title{Nuclear $\beta$-decay half-lives in the relativistic point-coupling model}

\author{Z. Y. Wang$^1$}
\author{Y. F. Niu$^{2,3}$}
\author{Z. M. Niu$^1$}\email{zmniu@ahu.edu.cn}
\author{J. Y. Guo$^1$}

\affiliation{$^1$School of Physics and Material Science, Anhui University,
             Hefei 230601, China}
\affiliation{$^2$Institute of Fluid Physics, China Academy of Engineering Physics,
             Mianyang 621900, China}
\affiliation{$^3$INFN, Sezione di Milano, via Celoria 16,
             I-20133 Milano, Italy}

\date{\today}

\begin{abstract}
The self-consistent proton-neutron quasiparticle random phase approximation approach is employed to calculate $\beta$-decay half-lives of neutron-rich even-even nuclei with $8\leqslant Z \leqslant 30$. A newly proposed nonlinear point-coupling effective interaction PC-PK1 is used in the calculations. It is found that the isoscalar proton-neutron pairing interaction can significantly reduce $\beta$-decay half-lives. With an isospin-dependent isoscalar proton-neutron pairing strength, our results well reproduce the experimental $\beta$-decay half-lives, although the pairing strength is not adjusted using the half-lives calculated in this study.
\end{abstract}

 \maketitle
The nuclear $\beta$ decay is an important property of nuclei, which can play an important role in nuclear physics and astrophysics. In nuclear physics, the investigation on $\beta$ decay can provide information on spin and isospin properties of nuclear effective interaction as well as nuclear properties such as masses~\cite{Lunney2003RMP}, shapes~\cite{Nacher2004PRL}, and energy levels~\cite{Tripathi2008PRL}. In astrophysics, about half of the nuclei heavier than Fe are synthesized by the rapid neutron-capture process ($r$ process). The $\beta$-decay rates of nuclei on the $r$-process path set the time scale of the $r$-process, hence determine the production of heavy elements in the universe~\cite{Burbidge1957RMP, Qian2007PRp, Langanke2003RMP}.

With the development of radioactive ion beam facilities, the measurement of nuclear $\beta$-decay half-lives has achieved great progress in recent years~\cite{Audi2012CPC}. For example, the measurements of the $\beta$-decay half-lives of neutron-rich nuclei in the vicinity of $N = 28$ and $N=50$ shell closures~\cite{Grevy2004PLB, Mazzocchi2013PRCa, Mazzocchi2013PRCb, Xu2014PRL}. On the theoretical side, apart from the macroscopic gross theory~\cite{Takahashi1975ADNDT}, two different microscopic approaches have been employed in the theoretical predictions of nuclear $\beta$-decay rates, i.e., the shell-model and the proton-neutron quasiparticle random phase approximation (QRPA).

The nuclear shell model can take into account the detailed structure of the $\beta$-strength function~\cite{Langanke2003RMP}, while the QRPA approach provides a systematic description of $\beta$-decay properties of heavy nuclei. The early phenomenological QRPA approaches for the $\beta$-decay half-lives rely on separably effective nucleon-nucleon interactions, such as the QRPA approach based on the deformed Nilsson+BCS formalism~\cite{Hirsch1992ADNDT} and on the finite-range droplet model (FRDM) with a folded Yukawa single-particle potential~\cite{Moller1997ADNDT}. Recently, the self-consistent QRPA approach has received more attention, which is generally believed to possess better extrapolation ability than the phenomenological QRPA.

The self-consistent QRPA approach has been employed in the calculation of nuclear $\beta$-decay half-lives based on the extended Thomas-Fermi plus Strutinsky integral (ETFSI) model~\cite{Borzov2000PRC}, the Skyrme-Hartree-Fock-Bogoliubov (SHFB) model~\cite{Engel1999PRC}, or the density functional of Fayans~\cite{Borzov1996ZPA}. On the other hand, the relativistic model has also received wide attention due to many successes in describing lots of nuclear phenomena~\cite{Meng2006PPNP, Vretenar2005PRp} and successful applications in astrophysics~\cite{Sun2008PRC, Niu2009PRC, Xu2013PRC}. In the relativistic framework, there are two widely used models, i.e., the finite-range meson-exchange and zero-range point-coupling models. In the meson-exchange model, the QRPA has been developed based on the relativistic Hartree-Bogoliubov (RHB) approach~\cite{Paar2004PRC} and employed in the calculation of $\beta$-decay half-lives of neutron-rich nuclei in the $N\approx 50$ and $N\approx 82$
regions~\cite{Niksic2005PRC, Marketin2007PRC}. Moreover, the self-consistent relativistic QRPA was also developed based on the relativistic Hartree-Fock-Bogoliubov (RHFB) theory~\cite{Liang2008PRL, Niu2013PLB} and it has been used to calculate $\beta$-decay half-lives of neutron-rich even-even nuclei with $20\leqslant Z\leqslant 50$. With an isospin-dependent $T=0$ proton-neutron pairing interaction, the known half-lives are well reproduced in the whole region of $20\leqslant Z\leqslant 50$¡£

In the point-coupling approach, the proton-neutron QRPA has been derived based on the effective Lagrangian with density-dependent couplings determined by chiral pion-nucleon dynamics~\cite{Finelli2007NPA}. In addition, the nonlinear point-coupling effective interactions have also received wide attention due to its successful description of lots of
nuclear phenomena~\cite{Burvenich2002PRC}. Recently, a new nonlinear point-coupling effective interaction, i.e., PC-PK1, was proposed, which well reproduces the properties of infinite nuclear matter and finite nuclei including the ground-state and low-lying excited states~\cite{Zhao2010PRC, Hua2012SCMPA, Meng2013FP, Mei2012PRC}. Based on the nonlinear point-coupling effective interaction PC-PK1, the self-consistent QRPA has been formulated and was employed to calculate the $\beta^+$/EC-decay half-lives of neutron-deficient Ar, Ca, Ti, Fe, Ni, Zn, Cd, and Sn isotopes~\cite{Niu2013PRCR}. The isospin-dependent pairing strength proposed in Ref.~\cite{Niu2013PLB} predicts a constant values of $V_0$ for these neutron-deficient nuclei and the calculations found that their half-lives are indeed reproduced well by an universal pairing strength~\cite{Niu2013PRCR}.

In this work, we will calculate the $\beta$-decay half-lives of neutron-rich nuclei with the newly proposed effective interaction PC-PK1. The isospin-dependent pairing strength is adopted but the parameters are directly taken from those in Refs.~\cite{Niu2013PLB, Niu2013PRCR}, i.e., not again adjusted with the known half-lives of calculated nuclei in this work.

The relativistic proton-neutron QRPA has been formulated in Ref.~\cite{Paar2004PRC} using the canonical single-nucleon basis of the RHB model for the meson-exchange effective interaction. Based on point-coupling relativistic Hartree-Bogoliubov theory, the self-consistent proton-neutron QRPA has also been established recently~\cite{Finelli2007NPA, Niu2013PRCR}. Similar to Ref.~\cite{Niu2013PRCR}, isovector-vector interaction, direct one pion interaction and the corresponding zero-range counter term are included in the QRPA particle-hole (p-h) residual interaction. The specific expressions for these terms can be found in Ref.~\cite{Niu2013PRCR}.

For the particle-particle (p-p) residual interaction, the isovector ($T = 1$) and isoscalar ($T = 0$) proton-neutron pairing interactions are included in the QRPA calculations. In the $T = 1$ p-p channel, we employ a phenomenological pairing interaction, i.e., the pairing part of the Gogny force
\begin{eqnarray}
        V_{T=1}(1,2)
   &=&  \sum_{i=1,2} e^{-[(\boldsymbol{r}_1-\boldsymbol{r}_2)/\mu_i]^2}\nonumber\\
   & &  (W_i + B_i P^\sigma - H_i P^\tau -M_i P^\sigma P^\tau),
\end{eqnarray}
with the parameter set D1S for $\mu_i, W_i, B_i, H_i$, and $M_i$~\cite{Berger1984NPA}. For the $T = 0$ proton-neutron pairing interaction in the QRPA calculation, we employ a similar interaction as in the Refs.~\cite{Engel1999PRC, Niksic2005PRC, Marketin2007PRC, Niu2013PLB, Niu2013PRCR}:
\begin{eqnarray}\label{Eq:Teq0}
    V_{T=0}(1,2)=-V_0 \sum_{j=1}^2 g_j e^{-[(\boldsymbol{r}_1-\boldsymbol{r}_2)/\mu_j]^2}
                 \hat{\prod}_{S=1,T=0},
\end{eqnarray}
with $\mu_1=1.2$ fm, $\mu_2=0.7$ fm, $g_1=1$, $g_2=-2$. The operator $\hat{\prod}_{S=1,T=0}$ projects onto states with $S=1$ and $T=0$.

The strength parameter $V_0$ in Eq.~(\ref{Eq:Teq0}) is usually determined by fitting to known half-lives of selected nucleus in each isotopic chain~\cite{Niksic2005PRC, Marketin2007PRC}.
However, very different values are found for different isotopic chains, which limits the prediction power of the model. For improving this dilemma, an isospin-dependent pairing strength $V_0$ was proposed in Ref.~\cite{Niu2013PLB}, i.e.,
\begin{eqnarray}\label{Eq:V02N}
  V_0 &=& V_L +\frac{V_D}{1+e^{a+b(N-Z)}},
\end{eqnarray}
where $V_L=134.0$ MeV, $V_D=121.1$ MeV, $a=8.5$, and $b=-0.4$ are obtained by fitting to the known half-lives of neutron-rich nuclei with the calculations of RHFB+QRPA approach. This isospin-dependent pairing strength predicts a constant values of $V_0$ for nuclei with $N-Z<5$, which is just the case for the neutron-deficient nuclei in the region $20\leqslant Z\leqslant 50$. For further testing the validity of this expression, the half-lives of neutron-deficient Ar, Ca, Ti, Fe, Ni, Zn, Cd, and Sn isotopes have been calculated with RHB+QRPA approach and PC-PK1 effective interaction~\cite{Niu2013PRCR}. It is found that their half-lives can be well reproduced by an universal pairing strength $V_0=175.0$ MeV, which indicates the validity of this expression of $V_0$. In this work, we will employ the RHB+QRPA approach and PC-PK1 effective interaction to calculate the half-lives of neutron-rich nuclei, so the value of $V_0=134.0$ MeV is replaced by $175.0$ MeV. Other parameters in Eq.~(\ref{Eq:V02N}) take the same values as in Ref.~\cite{Niu2013PLB}. Therefore, the parameters in Eq.~(\ref{Eq:V02N}) are not again adjusted with the known half-lives of calculated nuclei in this work.

In the allowed Gamow-Teller approximation, the $\beta$-decay half-life of an even-even nucleus is calculated with
\begin{eqnarray}\label{Eq:BetaDecayRate}
    T_{1/2}
  =\frac{D}{g_A^2 \sum_m B_m f(Z,E_m)}
\end{eqnarray}
where $D=6163.4\pm3.8$ s and $g_A=1$. The transition strength $B_m$ can be directly taken from the QRPA calculations. The integrated phase volume $f(Z,E_m)$ is the same as that in Ref.~\cite{Niu2013PLB}. The $\beta$-decay transition energy $E_m$ is calculated with
\begin{eqnarray}
    E_m = \Delta_{np} - E_{\textrm{QRPA}},
\end{eqnarray}
where $E_{\textrm{QRPA}}$ is the QRPA energy with respect to the ground state of the parent nucleus and corrected by the difference of the proton and neutron Fermi energies in the parent nucleus~\cite{Niu2013PLB, Niu2013PRCR} and $\Delta_{np}$ is the mass difference between neutron and proton. Since the emitted electron energy must be higher than its rest mass $m_e$, the final states must be those with excitation energies $ E_{\textrm{QRPA}} < \Delta_{np} - m_e = \Delta_{nH}$ ($\Delta_{nH}$ is the mass difference between the neutron and the hydrogen atom).

\begin{figure}[h]
  \includegraphics[width=6cm]{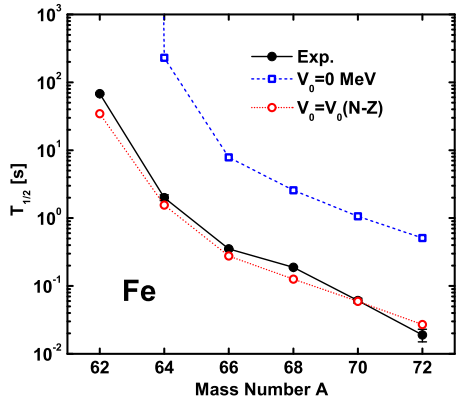}\\
  \caption{(Color online) Nuclear $\beta$-decay half-lives of Fe isotopes, calculated by the RHB+QRPA approach with the PC-PK1, compared to the available experimental data. The open squares and circles denote the calculations without and with the $T = 0$ pairing interaction, respectively.}\label{fig1}
\end{figure}
Fig.~\ref{fig1} shows the calculated half-lives of the Fe isotopes in comparison with the available experimental data. The results are calculated by RHB+QRPA model with the effective interaction PC-PK1. Obviously, the calculated half-lives are very sensitive to the $T = 0$ pairing
strength, which significantly reduce the calculated $\beta$-decay half-lives. Without the inclusion of this pairing channel, the calculated half-lives are systematically longer than the
experimental values. With the pairing strength parameter $V_0$ in Eq.~(\ref{Eq:V02N}), the half-lives of Fe isotopes are reproduced very accurately.

\begin{figure}[h]
  \includegraphics[width=6cm]{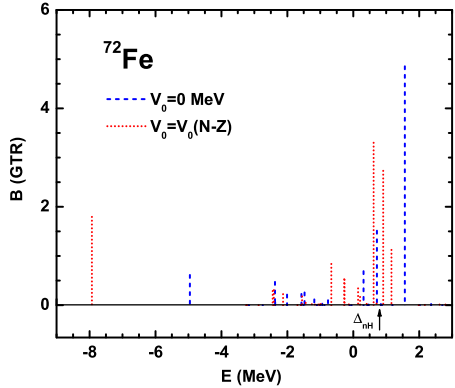}\\
  \caption{(Color online) The Gamow-Teller transition probabilities of $^{72}$Fe calculated without and with the $T = 0$ pairing interaction. The energy threshold $\Delta_{nH}$ for transitions which can contribute to nuclear $\beta$ decay is also denoted.}\label{fig2}
\end{figure}

The nuclear $\beta$-decay half-life is determined by the transition strength as well as the transition energy which decides the phase volume $f(Z,E_m)$. In order to illustrate the mechanism of the influence from $T = 0$ paring on the $\beta$-decay half-life, the Gamow-Teller transition strength distributions of $^{72}$Fe are shown in Fig.~\ref{fig2}. For the calculation without $T = 0$ pairing, the $\beta$-decay half-life of $^{72}$Fe is mainly determined by the transition at $E=-4.95$ MeV. This transition is dominated by the back spin-flip configuration $\nu 1f 5/2
\rightarrow \pi 1f 7/2$. Because both $\nu 1f 5/2$ and $\pi 1f 7/2$ orbits are partially occupied, the $T = 0$ pairing could give a remarkable contribution to the QRPA matrices related to the $\nu 1f 5/2 \rightarrow \pi 1f 7/2$ pair. When the attractive $T = 0$ pairing is included, the energy of transition dominated by the configuration $\nu 1f 5/2 \rightarrow \pi 1f 7/2$ is reduced to $E=-7.93$ MeV and simultaneously the transition strength is also increased. Therefore, the inclusion of $T = 0$ pairing can reduce the $\beta$-decay half-life.

\begin{figure}[h]
  \includegraphics[width=8cm]{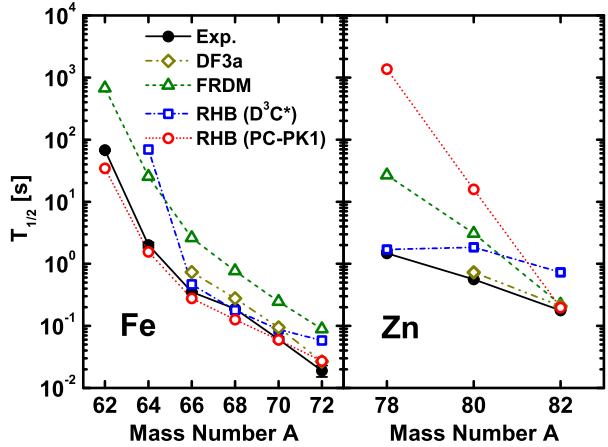}\\
  \caption{(Color online) Nuclear $\beta$-decay half-lives for Fe and Zn isotopes calculated by RHB+QRPA model with the effective interaction PC-PK1. For comparison, the experimental data and theoretical results obtained from the RHB+QRPA calculations with the effective interaction D$^3$C*~\cite{Marketin2007PRC}, the FRDM+QRPA calculations~\cite{Moller1997ADNDT}, as well as the continuum QRPA calculations~\cite{Borzov2003PRC} with the DF3a energy density functional~\cite{Tolokonnikov2010PAN} are also shown.}\label{fig3}
\end{figure}

For comparison with other theoretical results, nuclear $\beta$-decay half-lives for Fe and Zn isotopes are shown in Fig.~\ref{fig3}. Our calculations with PC-PK1 well reproduce the experimental data of Fe isotopes, including the recent experimental data of $^{72}$Fe. For Zn isotopes, the calculations with PC-PK1 overestimate the half-lives of $^{78, 80}$Zn, while it agrees well with the experimental half-life of $^{82}$Zn. Since $^{82}$Zn has crossed the shell closure of $N=50$, the single-neutron state $\nu 1g_{7/2}$ is partially occupied. This can lead to a transition dominated by the configuration $\nu 1g_{7/2} \rightarrow \pi 1g_{9/2}$, which is remarkably influenced by $T=0$ pairing. Therefore, the half-life of $^{82}$Zn is reduced and then well reproduce the experimental data. For the half-lives of $^{78, 80}$Zn, the influence of $T=0$ pairing is much weaker and recent studies have found that the tensor force~\cite{Minato2013PRL} and particle-vibration coupling~\cite{Niu2015PRL} can improve the overestimation of $\beta$-decay half-lives for nuclei with $N$ before the shell closure. For calculations with D$^3$C*, the results are close to experiment data for $^{66, 68, 70}$Fe with a $V_0=125$ MeV determined by fitting to half-lives of Fe isotopes, while an overestimation of half-lives of $^{64, 72}$Fe is observed. For reproducing half-lives of Zn isotopes, $V_0$ is refitted for the calculations with D$^3$C*. With an enhanced value of $V_0=300$ MeV, the half-life of $^{78}$Zn is well reproduced, while the half-lives of $^{80, 82}$Zn are still overestimated. Moreover, the results with FRDM+QRPA approach systematically overestimate the experimental $\beta$-decay half-lives for both Fe and Zn isotopes. This overestimation can be at least partially attributed to the neglect of $T=0$ paring~\cite{Engel1999PRC, Niu2013PLB}. In addition, by slightly modifying the spin-orbit terms in the DF3 functional, the modified functional DF3a has been developed and well describes the experimental data on spin-orbit splitting in magic and semimagic nuclei~\cite{Tolokonnikov2010PAN}. The available continuum QRPA calculations with DF3a agree well with the half-lives of both Fe and Zn isotopes.

\begin{figure}[h]
  \includegraphics[width=8.5cm]{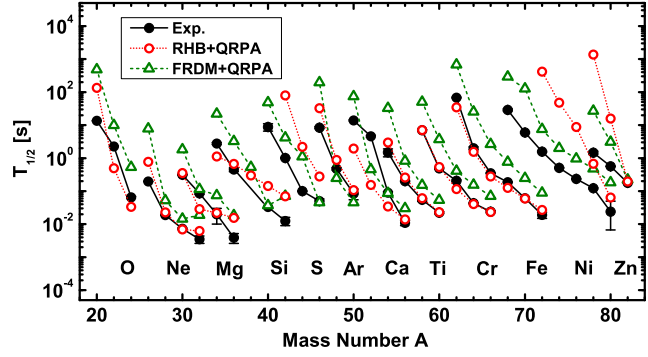}\\
  \caption{(Color online) Comparison of the calculated nuclear $\beta$-decay half-lives by RHB+QRPA model and effective interaction PC-PK1 with experimental data and FRDM+QRPA
  calculations for $Z=8-30$ even-even nuclei.}\label{fig4}
\end{figure}
Furthermore, with the isospin-dependent $T=0$ pairing strength in Eq.~(\ref{Eq:V02N}), we calculate the $\beta$-decay half-lives of neutron-rich even-even nuclei with $8\leqslant Z \leqslant 30$ by RHB+QRPA model and effective interaction PC-PK1. Fig.~\ref{fig4} displays the calculated results, experimental data, and FRDM+QRPA calculations. It is found that the FRDM+QRPA model systematically overestimates nuclear $\beta$-decay half-lives for most of nuclei in this region. However, the calculations with RHB+QRPA model generally agree well with the experimental data, except for S, Ni, and Zn isotopes, whose half-lives are systematically overestimated. The overestimation of half-lives for Ni isotopes can be understood as the same explanation for Zn isotopes. When the neutron number of Ni isotope crosses $N=50$ shell closure, the half-life deviation is remarkably reduced and even reproduce the recent experimental data of $^{80}$Ni~\cite{Xu2014PRL}. The overestimation of half-lives for Ni isotopes with neutron number before $N=50$ shell closure is a common problem in self-consistent relativistic QRPA calculations, which is also found in RHB+QRPA model with the meson-exchange effective interactions~\cite{Niksic2005PRC, Marketin2007PRC, Niu2013PLB}. By increasing the nucleon effective mass, half-lives of Ni isotopes can be reduced, hence become close to the experimental data. However, the increase of $m^*$ in relativistic model is limited by a realistic description of nuclear matter equation of state and ground-state properties of finite nuclei~\cite{Niksic2005PRC, Marketin2007PRC}. Therefore, some other effects in mean field framework, such as tensor force, and the effects beyond the mean field, such as the particle vibration coupling effect, should be investigated for better reproducing nuclear $\beta$-decay half-lives. For S isotopes, it was indicated that $^{40}$S and $^{42}$S are deformed $\gamma$-soft nuclei, while $^{44}$S exhibits shape mixing in the low-lying states~\cite{Sohler2002PRC, {Force2010PRL}}. Since we ignore deformation in present calculation, the discrepancy between theoretical results and experimental data may originate from the effect of deformation for S isotopes. With phenomenological QRPA model, it has been found that the inclusion of deformation indeed reduces the calculated half-lives~\cite{Grevy2004PLB}.

In summary, self-consistent proton-neutron quasiparticle random phase approximation approach is employed to calculate the nuclear $\beta$-decay half-lives with the relativistic nonlinear point-coupling effective interaction PC-PK1. It is found that the nuclear $\beta$-decay half-lives are sensitive to the isoscalar proton-neutron pairing interaction, which can significantly reduce the $\beta$-decay half-lives. With an isospin-dependent $T=0$ pairing strength, our results well reproduce the $\beta$-decay half-lives of neutron-rich even-even nuclei with $8\leqslant Z \leqslant 30$ except for Ni, Zn, and S isotopes, although the $T=0$ pairing strength is not adjusted with the half-lives calculated in this work. The underlying reasons for the overestimation of half-lives of Ni, Zn, and S isotopes are also discussed in detail.

This work was partly supported by the National Natural Science Foundation of China (Grants No. 11205004, No. 11305161, and No. 11175001), the 211 Project of Anhui University under Grant No. J01001319-J10113190081.



\end{document}